%% file: main.tex
\documentclass[conference]{IEEEtran}
\usepackage{cite}
\usepackage{amsmath,amssymb,amsfonts}
\usepackage{algorithmic}
\usepackage{graphicx}
\usepackage{textcomp}
\usepackage{xcolor}
\usepackage{balance}
\usepackage{xspace}
\usepackage{pifont}
\usepackage{subcaption}
\usepackage{booktabs}
\usepackage{colortbl}
\usepackage{multirow}
\usepackage{xurl}
\usepackage{array,ragged2e}
\usepackage{booktabs}
\usepackage{enumitem}
\usepackage{tablefootnote}

\def\BibTeX{{\rm B\kern-.05em{\sc i\kern-.025em b}\kern-.08em
    T\kern-.1667em\lower.7ex\hbox{E}\kern-.125emX}}

\newcommand{\ie}{\emph{i.e.,}\xspace}
\newcommand{\eg}{\emph{e.g.,}\xspace}

\begin{document}

\title{Defining a Reference Architecture for Edge Systems in Highly-Uncertain Environments}

\author{\IEEEauthorblockN{Kevin Pitstick, Marc Novakouski, Grace A. Lewis, Ipek Ozkaya}
\IEEEauthorblockA{\textit{Carnegie Mellon Software Engineering Institute}\\
Pittsburgh, PA USA \\
\{kapitstick, novakom, glewis, ozkaya\}@sei.cmu.edu}
}

\maketitle

\begin{abstract}
Increasing rate of progress in hardware and artificial intelligence (AI) solutions is enabling a range of software systems to be deployed
closer to their users, increasing application of edge software system paradigms. Edge systems support scenarios in which computation is placed closer to where data is generated and needed, and provide benefits such as reduced latency, bandwidth optimization, and higher resiliency and availability. Users who operate in highly-uncertain and resource-constrained environments, such as first responders, law enforcement, and soldiers, can greatly benefit from edge systems to support timelier decision making. Unfortunately, understanding how different architecture approaches for edge systems impact priority quality concerns is largely neglected by industry and research, yet crucial for national and local safety, optimal resource utilization, and timely decision making. Much of industry is focused on the hardware and networking aspects of edge systems, with very little attention to the software that enables edge capabilities. This paper presents our work to fill this gap, defining a reference architecture for edge systems in highly-uncertain environments, and showing  examples of how it has been implemented in practice. 
\end{abstract}


\input{introduction}
\input{high-uncertainty}

\input{software-arch}

\input{application}
\input{related-work}

\input{summary}

\section*{Acknowledgments}
This material is based upon work funded and supported by the Department of Defense under Contract No. FA8702-15-D-0002 with Carnegie Mellon University for the operation of the Software Engineering Institute, a federally funded research and development center (DM24-0158).

\bibliographystyle{IEEEtran}
\bibliography{references.bib}

\end{document}

%% file: introduction.tex
\section{Introduction}\label{sec:introduction}

Increasing pace of innovations in hardware and artificial intelligence (AI) and machine learning (ML) is enabling a range of software systems to be deployed closer to their users, a paradigm referred to as edge computing. Benefits typically attributed to edge computing include \cite{Shi2016}\cite{Yu2017}:
\begin{itemize}
\item \textbf {Reduced latency}: When data is processed close to its origin, it avoids round-trip time to remote servers or the cloud. This leads to faster response times, which enhances user experience.
\item \textbf{Bandwidth optimization}: When bandwidth is limited, edge software systems provide an avenue to pre-process, filter, and optimize what data is sent back to the cloud and when it should be sent.
\item \textbf{Higher resiliency and availability}: If connectivity to the cloud is unavailable, edge systems allow  processing to continue locally until connectivity is restored.
\item \textbf{Improved privacy and security}: Although often also considered a challenge or barrier, processing sensitive data locally avoids the risk of interception in network transmissions. In addition, data can be anonymized or cleaned before sharing, if needed.
\end{itemize}

For users that operate in highly-uncertain environments --- such as first responders, law enforcement, soldiers, rural medics, and any person performing tasks in the field where connectivity is unreliable --- an additional important benefit of edge computing is timelier decision making, supported by data processing closer to where it is generated. 
These benefits do not solely stem from the proximity of devices to users, but also depend on edge software systems being designed to maximize the benefits of edge devices and available networks, while managing necessary trade-offs. Edge systems supporting these users often need to access enterprise systems in the cloud, process locally-collected sensor data, and obtain wide-scale situational awareness. However, accessible networks may not guarantee reliable access to cloud resources, which is where this computation and data are typically available. While some of these cloud capabilities could be deployed locally to edge devices, these devices likely have limited resources that cannot properly execute these capabilities, especially for execution of AI and ML capabilities.

Unfortunately, while industry is focused on the hardware and networking aspects of edge computing, much less attention is being given to the software that enables edge  capabilities, despite the uncertainty inherent to the operating environment. 

This paper describes a reference architecture for edge software systems, targeted at managing and adapting to high levels of uncertainty in their operating environment. Section \ref{sec:high-uncertainty} describes the characteristics of highly-uncertain environments and the challenges they introduce. Section \ref{sec:software-arch} proposes a reference architecture for edge systems that operate in these environments. Examples of how portions of the architecture have been implemented in practice are described in Section \ref{sec:application}. Section \ref{sec:related-work} presents related work in this area and Section \ref{sec:summary} summarizes the paper and discusses next steps.

%% file: high-uncertainty.tex
\section{What Are Highly-Uncertain Environments?}\label{sec:high-uncertainty}

Highly-uncertain environments are characterized by uncertainty in several dimensions.
\begin{itemize}
\item{\textbf{Connectivity}: While connectivity to the cloud to access enterprise and other systems is not an issue in urban, industrial, or enterprise edge environments, it cannot be guaranteed in highly-uncertain edge environments, often referred to as DDIL (denied, degraded, intermittent and limited (bandwidth)). Decisions need to be made with limited network resources, regardless of the state of connections to greater resources.}
\item{\textbf{Computing resources}}: Because of connectivity challenges, teams operating in highly-uncertain environments only have reliable access to compute resources on edge devices that are with them, which limits the types of computing they can perform.
\item{\textbf{Security, safety, and stability of operational environment:} Depending on the situation, the security, safety, and stability of the operational environment will be affected by elements such as natural disaster, unrest, theft, and active attackers (physical and cyber). 
\item{\textbf{Operator Attention}}: In some of these environments, system operators experience cognitive load due to high levels of stress, which limits attention to user interfaces and how they can effectively interact with the system.}
\end{itemize}

These dimensions of uncertainty in edge environments, combined with resource limitations of edge devices, pose challenges to both the design of adaptation mechanisms and their execution during operations. The amount of uncertainty in the operating environment and available resources drive architecture design decisions for coping with the uncertainty. For example, a system that is expected to operate in an actively denied tactical network environment will be architected differently from one that operates in a network environment with spotty but high-bandwidth commercial cellular coverage. 

The next section proposes a reference architecture for edge systems that operate in environments that exhibit uncertainty across the stated dimensions. This architecture has been applied in three systems which we describe in Section \ref{sec:application}.

%% file: software-arch.tex
\section{Reference Architecture for Edge Systems in Highly-Uncertain Environments}\label{sec:software-arch}

Edge software systems are often part of larger distributed systems where there is interaction between cloud nodes and edge nodes, with edge nodes serving as intermediaries between proximate users and the cloud. Cloud-to-edge continuum (or edge-to-cloud continuum) is a term used to describe the interaction between cloud and edge nodes in which data and computation move between them as needed \cite{Milojicic2020}.

The main driver for how to meet requirements for edge systems that operate in highly-uncertain environments is the amount of uncertainty. To support the cloud-to-edge continuum, while managing and adapting to this uncertainty, edge systems will need to be: 
\begin{itemize}
    \item \textit{Network-aware}: Awareness of network conditions, such as availability and bandwidth, will drive decisions regarding when and what information to send.
    \item \textit{Location-aware}: Awareness of location, as well as other proximate devices and systems, drives decisions for computation distribution or delegation.
    \item \textit{Resource-aware}: Awareness of state of available resources, such as memory and storage, drives computation distribution and data storage decisions.
    \item \textit{Environment-aware}: Awareness of the environment via on-board sensors drives decisions related to computation distribution or lockdown, as well as switches between normal and degraded operation, for example.
    \item \textit{Secure by construction (with active monitoring}): Not having to rely on access to centralized authentication and authorization services enables access to local resources even when disconnected, and runtime adaptation of security postures in reaction to perceived threats. 
    \item \textit{Highly modular}: High modularity enables fine-grained computation distribution. 
\end{itemize}

\begin{figure*}[t!]
    \centering
    \includegraphics[width=16cm]{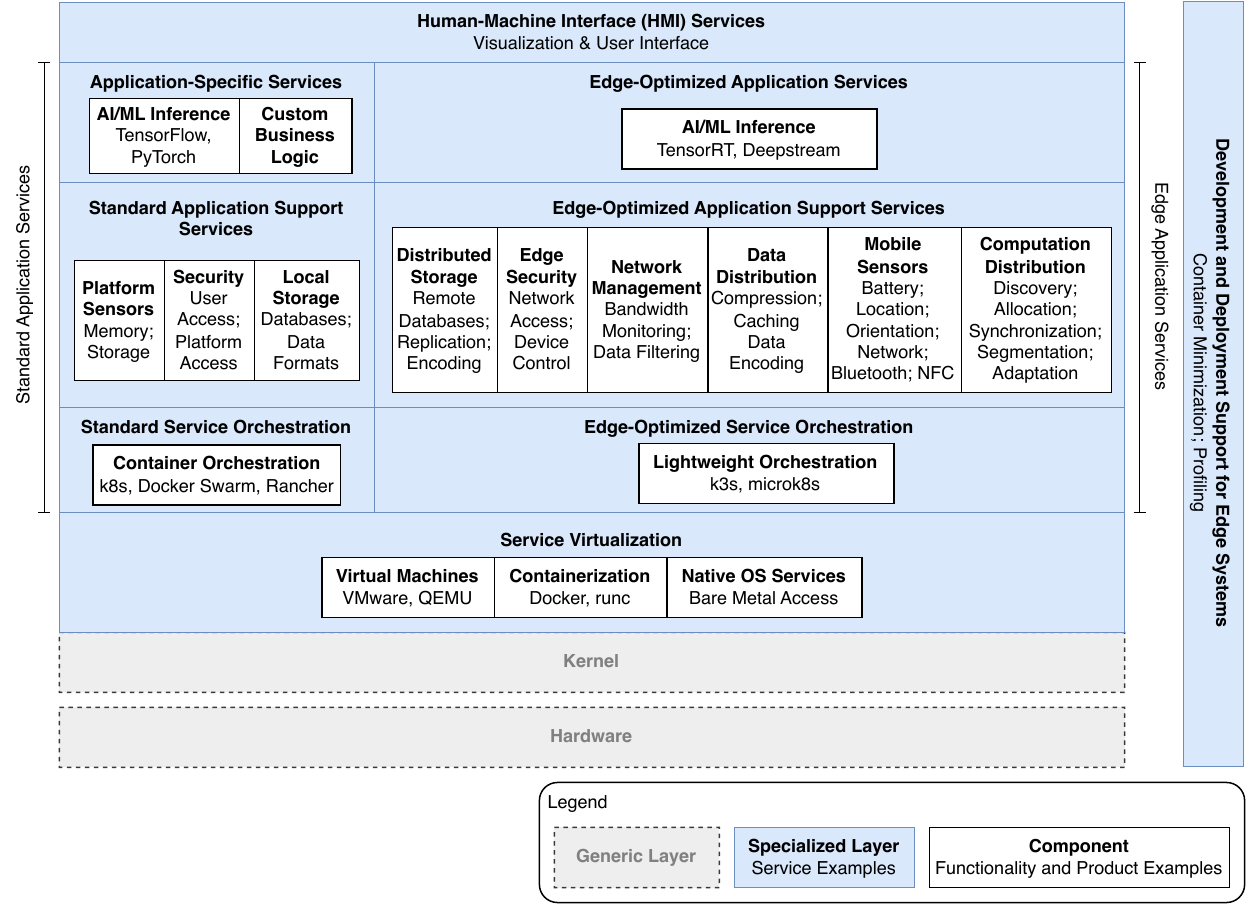}
    \caption{Proposed Reference Architecture for Edge Systems in Highly-Uncertain Environments — Layered View}
    \label{fig:edge-stack}
\end{figure*}

Figure \ref{fig:edge-stack} shows a proposed reference architecture for edge systems in highly-uncertain environments, that has emerged from our existing research in edge software systems, and refined through application in real systems, including those that are described in Section \ref{sec:application}. This architecture specifically calls out AI/ML-enabled capabilities as there is growing deployment of these capabilities at the edge, mainly for data processing and rapid decision making. AI/ML-enabled capabilities will in particular be challenged by computation resource constraints of edge devices due to their reliance on data and expensive computation needs \cite{Ding2022}. We now describe each of the layers of the architecture, from top to bottom.

\textbf{Human-Machine Interface (HMI) Services.} These services are all the visualization and user interface (UI) components that are supporting decision making in the field. HMI services must be simple and make use of advanced input methods, context-awareness, and automation to account for high cognitive load caused by operational environment conditions.

The left side of the middle layers in Figure \ref{fig:edge-stack} --- marked as Standard Application Services --- corresponds to more traditional services that apply to any edge software system. In many cases, because not all edge devices are extremely limited and are getting more powerful, they often do not even have to be edge-optimized: 
\begin{itemize}
    \item \textbf{Application-Specific Services.} These include custom business logic plus ML inference code for Edge AI/ML applications supported by traditional frameworks such as TensorFlow \cite{TensorFlow2024} and PyTorch \cite{PyTorch2024}.
    \item \textbf{Standard Application Support Services.} These include services for platform sensors for limited resources, such as memory and storage; local storage; and security for both user and platform access. 
    \item \textbf{Standard Service Orchestration.} Orchestration services are key for computation and data movement across cloud servers and edge devices and can be provided by commercial and open-source products such as Kubernetes (k8s) \cite{Kubernetes2024}, Docker Swarm \cite{Docker2024}, and Rancher \cite{Rancher2024}. 
\end{itemize} 

The right side of the middle layers in Figure \ref{fig:edge-stack} --- marked as Edge Application Services --- corresponds to edge-optimized services to deal with uncertainty and resource constraints. 

\textbf{Edge-Optimized Application Services}. Edge AI/ML applications on resource-constrained devices will need to include faster versions of traditional models, taking advantage of acceleration and optimization frameworks such as TensorRT \cite{TensorRT2024} and DeepStream \cite{DeepStream2024}. These frameworks are "hardware-aware", enabling them to make optimal acceleration decisions for the target device. In some cases, acceleration can reduce accuracy, leading to initial, sometimes coarse-grained, results. Even in these situations, these results allow users to react sooner, rather than wait for unreliable cloud resources to provide results, even if they provide more detailed, accurate, and finer-grained answers.

\textbf{Edge-Optimized Application Support Services}. These are the main services that collect information about the environment, optimize resource usage, and adapt to operational conditions.
\begin{itemize}
\item \textit{Distributed Storage}  
services support distributed data storage solutions across cloud and edge nodes, including support for remote databases, replication for reduced latency and disconnected operations support, and optimized encoding for reduced storage.
\item \textit{Edge Security} services are used for network access and device control, including decentralized methods for authentication and authorization to account for operation in DDIL environments \cite{Echeverria2019}.
\item \textit{Network Management} services manage available network bandwidth, including continuous bandwidth monitoring for adaptation triggers and data filtering to optimize bandwidth usage.
\item \textit{Data Distribution} services are used to optimally distribute data between cloud and edge, including compression, caching, and data encoding, such as protocol buffers \cite{ProtocolBuffers2024}, DTN bundles \cite{Scott2007}, and EXI encoding for XML \cite{W3C2014}.
\item \textit{Mobile Sensor} services manage any available on-board sensors that will trigger adaptations in response to changing conditions or available resources, such as battery, location, orientation, network, Bluetooth, and NFC (near-field communications). Sensor outputs can also be used to trigger UI changes to adapt to operational conditions.
\item \textit{Computation Distribution} 
services are used to adapt and distribute computation in response to changing conditions and available resources, including discovery,  allocation, synchronization, segmentation, and adaptation.
\end{itemize}

\textbf{Edge-Optimized Service Orchestration.} These are orchestration services for constrained edge devices provided by commercial and open-source products such as k3s \cite{K3S2024} and MicroK8s \cite{MicroK8s2024}.

\textbf{Service Virtualization.} Given the cloud-to-edge continuum and edge device heterogeneity, service virtualization is key for computation distribution from cloud to edge, but also between edge devices. Services in this layer include support for virtual machines, containers, and native OS services.

\textbf{Development and Deployment Support for Edge Systems}. Cross cutting to all layers are services for development and deployment of edge systems in highly-uncertain environments, such as container minimization to address limited bandwidth and computing resources (\eg SlimToolkit \cite{Slim2024}), and profiling to understand if an application can effectively perform on edge devices (\eg htop \cite{htop2024}).

%% file: application.tex
\section{Examples of Application}\label{sec:application}

What follows are three examples of how we have implemented different layers and components of the reference architecture in practice. We elaborate on architecture elements and decisions that are relevant to managing high uncertainty at the edge. Edge system names are omitted due to confidentiality. 

\subsection{System A}
This system is a computing environment designed to provide situational awareness to personnel operating at the edge, such as first responders. The system is designed to support both design-time and runtime integration of multiple edge subsystems to support a mix of planned and unplanned field tasks, such as search and rescue, and medical triage. Examples of subsystems include user interfaces, visualizations, infrastructure management, task tracking, and situational awareness. Uncertainty in this system comes from the unknown availability of compute nodes, requiring shifting of computation between nodes. The architecture drivers for this system include:
\begin{itemize}
    \item Interoperability: The system needs a clearly defined messaging interface to allow subsystems to easily exchange data.
    \item Availability/Survivability: The system contains multiple compute nodes and needs to continue operating even if only one compute node is still operational.
    \item Adaptability: With the high level of mission uncertainty, the system needs to be able to rapidly shift which services are running based on the user's intent and direction.
    \item Integratability: The system needs to make it easy for system operators and developers to integrate other edge subsystems rapidly and easily, such as a new sensor module required for the mission.
    \item Security: The system needs to  communicate with other systems securely, without broadcasting sensitive data in plain-text.
\end{itemize}
Architecture element decisions in the \textit{Edge-Optimized Application Support Services} layer are the most relevant to meet these drivers and include the following:
\begin{itemize}
    \item Distributed Storage: As availability and survivability are key concerns, the system needs to continue operating when only a single node remains operational. To achieve this, Postgres BDR (Bi-Directional Replication) \cite{PostgresBDR2024} was utilized to sync the database state of the system across all nodes, achieving multi-master replication.
    \item Edge Security: The system contains dedicated radios with built-in encryption to allow it to securely communicate with off-board systems.
   \item Data Distribution: To meet interoperability and resource-constraint demands, Google's Protocol Buffers \cite{ProtocolBuffers2024} are utilized for encoding of all data being transmitted. Data is distributed using ZeroMQ \cite{ZeroMQ2024} directly between services, as opposed to centralized or proxy-based distribution, to ensure that distribution is timely without a single point of failure. Distributed middleware achieves optimized data flows through the use of direct connections between communicating services, mediated by a special service that ensures seamless and automated connection management.
   \item Mobile Sensors: Mobile sensors are leveraged to gather data related to situational awareness (\eg GPS, audio, system power, health status (\eg ECG, pulse ox)) and inform users of changing conditions.
\end{itemize}

Other key decisions include those related to the \textit{Standard Service Orchestration} and \textit{HMI Services} layers of Figure \ref{fig:edge-stack}.  
\begin{itemize}
    \item Containerization: The use of Docker containers is a key decision to promote integratability of diverse applications by multiple vendors, on the same compute platforms, in a reasonable timescale. Containers also promote adaptability because they are self-contained units of compute that can be shifted between compute nodes.
    \item Lightweight Orchestration: A custom lightweight orchestration component is built on top of Docker using open-source standard services and industry standard container management interfaces to ensure that the right services are running at the right time, and that services to support any field task are ready to execute when needed (note that this was before the emergence of products such as k8s and k3s), promoting adaptability. Because the edge platforms of System A are not extremely resource-constrained, redundant services are deployed on all edge nodes, promoting availability.
    \item HMI Services: Contextual clues, such as alerts and highlights, are presented using augmented reality and near-eye displays to prevent cognitive load, combined with simple preset button controls for fast, intuitive, and customizable user interaction.
        
  \end{itemize}

\subsection{System B}
This system is a distributed ML-enabled pipeline for natural language translation. Each element of the pipeline, such as speech-to-text, language translation, and text-to-speech, is a separate service hosted on a separate edge device. A key requirement for this system is data distribution in a dynamic environment with edge devices connecting and disconnecting at various intervals, either intentionally or as the result of failure. This is the largest source of uncertainty in the system and requires adaptability and flexibility for building an end-to-end pipeline during mission time based on operator requests. The architecture drivers for this system include:
\begin{itemize}
    \item Adaptability/Flexibility: The system needs to adapt which ML elements are running on each device based on device availability and system load.
    \item Interoperability: The system needs a clear messaging interface for heterogeneous devices to exchange messages.
    \item Availability: The system needs to continue processing the ML pipeline while operating on a network with uncertain connectivity between devices.
    \item Performance: The system needs to run ML services, which have high computational requirements.
\end{itemize}
Critical design decisions center around \textit{Edge-Optimized Application Services}, \textit{Edge-Optimized Application Support Services}, and \textit{Edge-Optimized Service Orchestration} layers. 
\begin{itemize}
    \item Edge-Optimized Application Services: The system utilizes both custom and off-the-shelf lightweight ML models for the pipeline's language tasks, leveraging edge-optimized streaming frameworks like NVIDIA Deepstream \cite{DeepStream2024}. To achieve the required edge performance, NVIDIA Jetson devices are deployed, which have on-board GPUs for ML inference while keeping a low size/power footprint.
    \item Computation Distribution: Each device is configured to load one or more containers that compute a single stage of the ML pipeline (\eg speech-to-text), allowing computation to be distributed across all available edge devices.
    \item Data Distribution: Similar to System A, all data is encoded using Protocol Buffers \cite{ProtocolBuffers2024} to aid in interoperability between the services running on different devices. However, rather than using ZeroMQ \cite{ZeroMQ2024}, this system utilizes ROS2 \cite{ROS2024} (and thus Fast DDS \cite{FastDDS2024}) as the messaging middleware. Fast DDS seamlessly provides adaptability during network disconnect/reconnect sequences in addition to providing quality of service (QoS) profiles to handle DDIL network conditions.
    \item Containerization: Docker containers are utilized to encapsulate each processing component of the ML pipeline and its dependencies, allowing them to run on any of the compatible edge devices through the use of the NVIDIA Container Toolkit.
    \item Lightweight Orchestration: The orchestrator assigns pipeline elements to edge devices as they became available. This promotes adaptability as the system can quickly reorganize for a new mission, as well as flexibility because the service layout for a given mission can be rearranged as necessary as available resources fluctuate up and down. In addition, the orchestrator's job is to achieve maximum performance given the current state of computational devices available, rebalancing which ML elements are running on each device based on queue sizes at each stage of the pipeline.
\end{itemize}

\subsection{System C}
This smaller system is a communications gateway that translates between different protocols to enable system interoperability at the edge. Uncertainty in this system comes from the dynamic mission environment, as well as the wide range of known and unknown systems that may be targeted for integration. Architecture drivers for this system include:
\begin{itemize}
    \item Interoperability: The system needs to enable communications between heterogeneous systems, developed by different organizations using a variety of platforms, languages, and tools.
    \item Portability: The system needs to be easily and quickly deployed on a variety of edge devices.
    \item Performance: Maximizing throughput is critical to successful operation, due to expected network challenges of the environment. 
\end{itemize}
The most relevant architecture element decisions to support these drivers are in \textit{Data Distribution} in the \textit{Edge-Optimized Application Support Services} layer:
\begin{itemize}
    \item High performance, low-level services (\ie C libraries) for network-efficient message encoding (\eg variable bit encoding schemes such as Protobuf) are implemented to maximize system performance at the edge. 
    \item Different edge-optimized data distribution services (\eg different versions of DDS) are used to optimize data distribution across a variety of network setups, with explicit built-in interfaces to allow for switching to other DDS services. This promotes interoperability with systems and networks that have already made investments in data distribution technologies.
\end{itemize}
Similar to Systems A and B, a key decision related to the \textit{Service Virtualization} layer of Figure \ref{fig:edge-stack} is \textit{Containerization}. The use of Docker containers enables maximum portability in heterogeneous environments (\eg different operating systems and/or hardware) and integration with heterogeneous edge devices, therefore promoting interoperability. External interfaces to containers use standard and near-universal socket-level communications (\eg TCP and UDP), which enables easy integration with other systems, further promoting interoperability.

%% file: related-work.tex
\section{Related Work}\label{sec:related-work}

A key aspect of operating in highly-uncertain environments is adapting to change. While there is a large body of work studying adaptive and self-adaptive software systems \cite{Salehie2009}, outcomes are often not mature enough to be deployed in these environments \cite{Szabo2020}. Moreover, these outcomes lack an explicit focus on enumerating sources of uncertainty, eliciting ranges of variability, and mapping design approaches accordingly. 

In the context of edge system architecture, there is work defining general reference architectures for edge systems \cite{Sitton2019}\cite{Prokhorenko2020}, and specific system types, such as industrial edge systems \cite{Willner2020} and IoT systems \cite{Fernandez2018}. However, and consistent with the motivation for this work, the focus is on hardware, networks, and the definition of layers across cloud, edge, and end devices (\eg sensors, mobile devices), and not on the edge software system that runs on edge devices.

As far as operation in highly-uncertain environments, there is recognition of the challenges of operating in hostile and contested environments \cite{Satyanarayanan2013}, especially in the context of military systems \cite{Singh2017}\cite{Fraga2022} and emergency management \cite{DOro2019}\cite{Wu2017}. However, the focus of this work is also mostly on hardware, network, and infrastructure, rather than software and software architecture.

Finally, industry has made huge advances in edge computing in recent years, including single-board computers (\eg NVIDIA Jetson), cloud infrastructure for the edge (\eg AWS Snowball Edge), and processors (\eg Intel Xeon). The communications industry is also enabling edge computing through widespread 5G availability and mobile access points. However, industry solutions remain very hardware focused, and vertical solutions are typically not related to highly-uncertain environments (\ie transportation, retail, and manufacturing, as opposed to defense and public safety) \cite{Chabas2018}. 

This work fills the software architecture gap for edge systems in highly-uncertain environments, a domain largely neglected by industry and research.

%% file: summary.tex
\section{Summary and Next Steps}\label{sec:summary}

This paper describes a layered view of a reference architecture for edge software systems that operate in highly-uncertain environments, which has been developed while addressing architecture challenges in varying contexts, as shown in Section \ref{sec:application}. As advances in hardware, communications, and AI and ML capabilities continue, there will be an increasing need for defining better software architectures for edge systems. Challenges that will continue to have an effect on how edge software systems are architected include:
\begin{itemize}
    \item Adapting computation across multiple nodes as resources change
    \item Leveraging AI/ML for real-time decision making
    \item Managing and processing the immense amount of data collected at the edge
    \item Dealing with compatibility and interoperability as hardware diversity increases
    \item Securing systems in highly-uncertain environments.
\end{itemize}

Our next steps include continuing to refine, define, and implement portions of the reference architecture, and to describe it from other perspectives as we apply it to more edge software systems operating in highly-uncertain environments.